# Electrical and thermal transport properties of kagome metals $A$V$_3$Sb$_5$ ($A$=K, Rb, Cs)

Xin-Run Mi[1], Kun-Ya Yang[1], Yu-Han Gan[1], Long Zhang[1], Ai-Feng Wang[1],

Yi-Sheng Chai[1], Xiao-Yuan Zhou [1*], Ming-Quan He[1*]

[1] Low Temperature Physics Lab, College of Physics & Center of Quantum Materials

and Devices, Chongqing University, Chongqing 401331, China

*Corresponding authors:

xiaoyuan2013@cqu.edu.cn (Xiao-Yuan Zhou)

mingquan.he@cqu.edu.cn (Ming-Quan He)

**Abstract**

The interplay between lattice geometry, band topology and electronic correlations in the newly discovered kagome compounds $A$V$_3$Sb$_5$ ($A$=K, Rb, Cs) makes this family a novel playground to investigate emergent quantum phenomena, such as unconventional superconductivity, chiral charge density wave and electronic nematicity. These exotic quantum phases naturally leave nontrivial fingerprints in transport properties of $A$V$_3$Sb$_5$, both in electrical and thermal channels, which are prominent probes to uncover the underlying mechanisms. In this brief review, we highlight the unusual electrical and thermal transport properties observed in the unconventional charge ordered state of $A$V$_3$Sb$_5$, including giant anomalous Hall, anomalous Nernst, ambipolar Nernst and anomalous thermal Hall effects. Connections of these anomalous transport properties to time-reversal symmetry breaking, topological and multiband fermiology, as well as electronic nematicity, are also discussed. Finally, a perspective together with challenges of this rapid growing field are given.

# 1 Introduction

Emergent quantum phases of matter, such as high-temperature superconductivity, fractional quantum Hall effect, often arise in quantum materials carrying entangled lattice, charge, spin and orbital degrees of freedom [1,2]. The kagome lattice as proposed by Syôzi back in 1951 [3], is one of the most studied systems for realizing various emergent quantum states. The kagome lattice is consisted of corner-sharing triangles, giving rise to various nontrivial features in electronic structure, such as linearly dispersing Dirac cones, van Hove singularities and a dispersionless flat band [4–8]. The flat band naturally features large electron effective mass with strong correlation effects. Combined with the special lattice geometry, the kagome lattice is an ideal platform to study the interplay of geometrical frustrations, strong correlations and band topology [4–8]. At the theoretical level, a variety of quantum phases, ranging from quantum spin liquids, Dirac and Weyl fermions, fractional Chern insulators, to unconventional superconductivity, etc. could be realized in kagome lattice-based systems [9–13] Experimentally, a few kagome materials, such as quantum spin liquid candidate $ZnCu_3(OH)Cl_2$ [14,15], magnetic Weyl semimetals $Co_3Sn_2S_2$ [16] and $Mn_3Sn$ [17], Dirac Metals $Fe_3Sn_2$ [18] and $FeSn$ [19], have been identified. Investigation of novel quantum phases of matter in these compounds, as well as exploring new kagome materials, are of great interest in condensed matter physics and material science.

In 2019, Ortiz et al. [20] discovered a new group of kagome metals $AV_3Sb_5$ ($A$=K, Rb, Cs), which offers another prominent material system to study intriguing physics arising from the kagome lattice. In $AV_3Sb_5$, the V atoms are coordinated nicely in a kagome network [see Fig. 1a]. The resultant electronic structure hosts Dirac-like dispersions, $\mathbb{Z}_2$ topological characters and multiple van Hove singularities in the vicinity of Fermi level [21]. Along with these nontrivial electronic features, a variety of emergent phases arise in $AV_3Sb_5$. Most notably, all three members in the $AV_3Sb_5$ family undergo a charge density wave (CDW) transition at $T_{CDW}$=78-103 K, and a superconducting transition at $T_c$=0.9-2.5 K [21–23]. More importantly, both the charge and superconducting orders likely carry unconventional characters, which are intimately coupled to the unique lattice geometry and electronic structure. The CDW state probably has chirality that breaks time-reversal symmetry, accompanied by an exceptionally large anomalous Hall effect (AHE) [5,8,24–28]. While the nature of the superconducting state is still under debate, a few unusual features, such as gap nodes, Majorana bound states, nematicity and roton-like pair density wave, have been reported [29–33]. Additionally, CDW competes with superconductivity in a complex way when tuned by hydrostatic pressures [34–40]. The kagome metals $A$V$_3$Sb$_5$ have become an outstanding playground to uncover the intricate interplay of intertwined orders.

Among various extensive studies, electrical and thermal transport measurements have played important roles, including the observations of large AHE, electronic nematicity, anomalous Nernst effect (ANE), ambipolar Nernst effect and anomalous thermal Hall effect (ATHE). In this review, we briefly summarize these unusual electrical and thermal transport properties of $A$V$_3$Sb$_5$. We start with an introduction to crystal and electronic structures, followed by description of CDW and superconductivity of $A$V$_3$Sb$_5$. Then we focus on the electrical and thermal transport properties of $A$V$_3$Sb$_5$ in the CDW state. Finally, discussions on open questions and an outlook are presented.

## 2 Crystal and electronic structures

Fig. **1** presents the crystal structure and electronic fermiology of the kagome metals $A$V$_3$Sb$_5$. All three members in $A$V$_3$Sb$_5$, i.e., KV$_3$Sb$_5$, RbV$_3$Sb$_5$ and CsV$_3$Sb$_5$ share similar structures with moderate differences due to different atomic sizes of the alkali-$A$ atoms. As shown in Fig. 1a, the $A$V$_3$Sb$_5$ compounds crystallize in a hexagonal structure (*P6/mmm*). At room temperature, a perfect kagome lattice is formed by V atoms in the V-Sb layer, which is sandwiched by triangular alkali-$A$ layers. There are two types of Sb atoms sitting at different positions. The Sb1 atoms are arranged in a honeycomb fashion, positioning above and below the kagome layer. The Sb2 atoms locate at the centers of hexagons in the V-kagome lattice, forming a triangular pattern. The $A$V$_3$Sb$_5$ compounds are thus layered materials and the essential parts are the V-kagome layers. As shown in Fig. 1b, c, a kink is observed at $T_{\text{CDW}}$=94 K in resistivity, and zero resistivity is found below $T_c$=2.5 K for CsV$_3$Sb$_5$. These features are resulted from the CDW and superconducting instabilities, respectively. The large ratio $\rho_c/\rho_{ab} \sim 600$ between out-of-plane resistivity ($\rho_c$) and in-plane resistivity ($\rho_{ab}$) implies a quasi-two-dimensional nature.

In Fig. 1d, the electronic band structure obtained by density functional theory (DFT) calculations is shown for CsV$_3$Sb$_5$ [21]. Similar features are found in KV$_3$Sb$_5$ and RbV$_3$Sb$_5$. The calculated results are consistent with the experimentally determined structures seen in angle-resolved photoemission spectroscopy (ARPES) measurements [21]. Clearly, the $AV_3Sb_5$ compounds are multiband metals with a few bands intersecting with the Fermi level. As shown in Fig. 1e [41], in the Brillouin zone center, a circular electron pocket ($\alpha$) composed of Sb $p_z$ orbitals is located around the $\Gamma$ point. Near the zone boundaries, tiny pockets involving the $d$ orbitals of V atoms are found, including triangular-shaped electron ($\delta$) and hole ($\gamma$) bands around the $K$ point, a small hole pocket near the $M$ point and a tiny electron pocket around the $L$ point. Moreover, typical features of the kagome lattice are seen. These include multiple Dirac-like dispersive bands and van Hove singularities siting near the $M$ point close to the Fermi energy. Fermi surface nesting between the van Hove singularities could lead to the CDW instability. The band topology is further enriched by the nontrivial topological $\mathbb{Z}_2$ invariant obtained for the bands crossing the Fermi level. In particular, the parity index switches sign between different bands at the time-reversal invariant $M$ point (see the right panel in Fig. 1e), leading to band inversion and topological protected surface states around the $M$ point, as confirmed by ARPES experiments (see Fig. 1f) [42]. These nontrivial features in electronic fermiology have profound impacts on physical properties, which we will discuss in detail below.

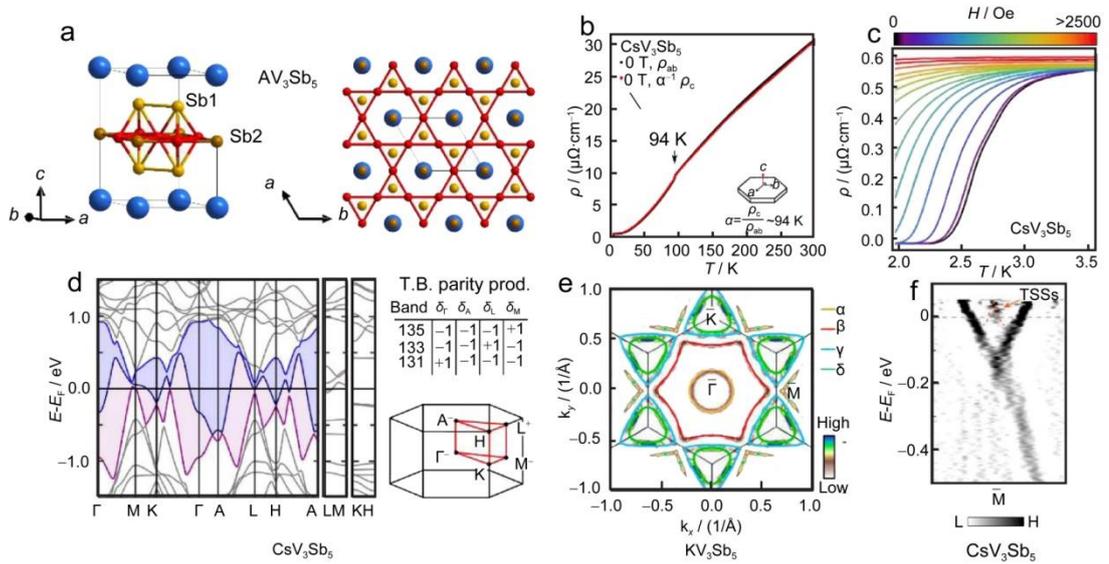

**Fig. 1 a** Crystal and electronic band structures of $AV_3Sb_5$. Side view (left) and top view (right) of the crystal structure (*P6/mmm* space group). A perfect kagome lattice is formed by V atoms in the V-Sb2 layer. **b** and **c** Signatures of CDW and superconducting transitions of $CsV_3Sb_5$ seen at 94 and 2.5 K in electrical resistivity measurements. **d** Electronic band structure of $CsV_3Sb_5$ obtained by DFT calculations. **e** Fermi surfaces of $KV_3Sb_5$ resolved by ARPES experiments in comparison with those calculated by DFT (solid lines). **f** Topological surfaces states (TSSs, red dash lines) detected by ARPES in $CsV_3Sb_5$ at the *M* point near the Fermi level. **b, c** and **d** Adapted with permission from Ref. [21]. Copyright 2020 American Physical Society (APS). **e** Data from Ref. [41]. **f** Reproduced with permission from Ref. [42]. Copyright 2021 Elsevier.

# 3 Charge density wave

Upon cooling from room temperature, the leading symmetry breaking experienced by all three members in $AV_3Sb_5$ is a charge density wave instability, which occurs at $T_{CDW}$=78, 103 and 94 K for $KV_3Sb_5$, $RbV_3Sb_5$ and $CsV_3Sb_5$, respectively [21–23]. The charge order is modulated in a 2×2 periodicity within the kagome *ab* plane (see Fig. 2a), as revealed by various techniques, including X-ray diffraction [22,43,44], scanning tunneling microscopy (STM) [24,30,33,45–47] and ARPES [48–51]. The out-of-plane modulation of the charge order is still under debate. Two types of modulation including 2×2×2 [24,28,30,43,46,52,53] and 2×2×4 [44] have been reported. And these two forms may even coexist [54,55] or appear in different temperature intervals [56]. Additionally, a 4*a* unidirectional stripe charge order has also been identified, which likely originates from surface effects [28,33,45,46,57].

The CDW instability necessarily leads to energy gap opening and reduction of density of states near the Fermi energy. Due to the multiband nature, the CDW gaps depend strongly on the Fermi pockets and momentum [41]. As shown in Fig. 2b, the charge ordering is nearly independent of the Sb-$p_z$ orbital-derived bands, and no gap is seen for the electron pocket around the $\varGamma$ point. The CDW gaps mainly appear in the V-*d* orbitals-derived pockets at the zone boundaries, and are highly anisotropic. This indicates that the CDW mainly affects the V-kagome net, as we will see in the lattice deformations below. The largest gap appears at the nested vectors of the Fermi surface, likely suggesting an electronic driven CDW instability [5]. This scenario is also supported by hard-X-ray diffraction experiments, which do not find the acoustic phonon anomaly that typically appears in electron-phonon coupling-driven CDW [43]. On the other hand, sizable electron-phonon coupling has been reported in ARPES, Raman scattering, neutron scattering and optical spectroscopy measurements [58–60]. The electron-phonon coupling may even dominate in the CDW transition, as suggested theoretically [61]. Thus, the exact nature of the CDW is still elusive, and further investigations are desired to uncover the leading driving force.

The interplay of charge and lattice degrees of freedom also leads to moderate lattice distortions in the CDW phase [44,62]. As displayed in Fig. 2c, unstable phonon modes near $M$ and $L$ points can lead to two possible types of in-plane lattice deformation, i.e. Star of David (SD) and Inverse Star of David (ISD), in accordance with the observed in-plane 2×2 superlattice modulation [62]. It is suggested by DFT calculations that the ISD structure is energetically more stable, compared to the SD case [61,62]. Experimentally, various structures including pure SD [63], single ISD [55,64,65] and coexistence of SD/ISD [44,54] have been found. Stacking of SD/SD, or ISD/ISD, or SD/ISD with an inter-layer phase shift of $\pi$ results in a three-dimensional 2×2×2 periodicity. These kinds of inter-layer stacking also reduce the $C_6$ rotational symmetry to the $C_2$ symmetry, as seen by X-ray diffraction, STM, electrical transport and optical Kerr experiments [30–32,47,55,56,66,67].

The most notable feature of the CDW is possible existence of time-reversal symmetry breaking in the absence of any long-range magnetic orders. This peculiar property has been evidenced by various studies. As shown in Fig. 2d, the intensities detected by STM at the CDW ordering vectors are inequivalent, displaying a chirality feature which can be tuned by an external magnetic field [24,45,46]. The existence of time-reversal symmetry breaking is further supported by the observation of internal magnetic fields in the CDW state using muon spin relaxation ($\mu$SR) (see Fig. 2e) and polar Kerr measurements [27,28,66,68]. These observations are also in line with the giant AHE which appears concomitantly with the CDW transition [25,26,69]. We will discuss the AHE in more detail later. Motivated by these experimental findings, theoretical studies have proposed a chiral flux scenario, which naturally breaks time-reversal symmetry, as shown in Fig. 2f [70,71]. This chiral flux phase is reminiscent of the orbital currents model of quantum anomalous Hall effect proposed by Haldane [72], and the loop currents model of high-temperature cuprate superconductors suggested by Simon and Varma [73]. The theoretically proposed chiral flux phase for $A$V$_3$Sb$_5$ is consistent with the experimentally observed in-plane 2×2 charge modulation. However, no signature of this chiral flux phase has been detected in CsV$_3$Sb$_5$ using spin-polarized STM experiments [74]. In addition, another STM study suggests that the CDW order of KV$_3$Sb$_5$ is not sensitive to the direction of external magnetic fields [47].

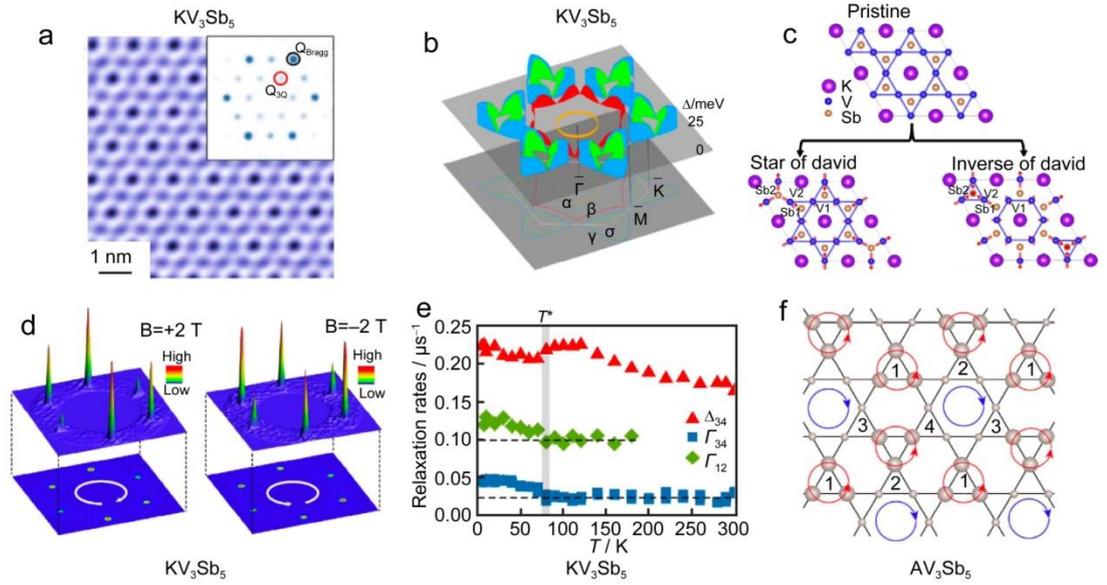

**Fig. 2** Charge density wave of $AV_3Sb_5$. **a** The 2×2 superlattice charge modulation resolved by STM on the Sb surface of $KV_3Sb_5$. **b** Momentum-dependent CDW gap seen by ARPES in $KV_3Sb_5$. **c** Theoretically proposed two types of lattice distortions induced by CDW. **d** Charge modulation peaks observed at the CDW vectors in $KV_3Sb_5$. The intensities of these peaks show chirality that is tunable by external magnetic fields. **e** Temperature-dependent muon relaxion rates of $KV_3Sb_5$, which increase gradually below the CDW transition temperature, indicating the existence of internal magnetic fields in the CDW sate. **f** The chiral flux model proposed theoretically. **a** and **d** Adapted with permission from Ref. [24]. Copyright 2021 Springer Nature. **b** Data from Ref. [41]. **c** Reproduced with permission from Ref. [62]. Copyright 2021 APS. **e** Adapted with permission from Ref. [28]. Copyright 2022 Springer Nature. **f** Data from Ref. [70].

# 4 Superconductivity

Intriguingly, superconductivity emerges inside the CDW phase at $T_c$=0.9, 0.9 and 2.5 K for $KV_3Sb_5$, $RbV_3Sb_5$ and $CsV_3Sb_5$, respectively [21–23]. Despite the low $T_c$ values, it has been theoretically suggested that electron-phonon coupling is too weak to be responsible for the observed $T_c$, possibly pointing to an unconventional superconducting nature [62]. The unconventional superconductivity is further evidenced by residual thermal conductivity $\kappa_0(T\to 0$ K$)$, which shows $d$-wave-like magnetic field dependence and nodal superconductivity features, as shown in Fig. 3a for $CsV_3Sb_5$ [29]. In the vortex state as presented in Fig. 3b, signatures of Majorana zero modes have been identified in the vortex core by STM measurements, suggesting a nontrivial superconducting state [30]. Moreover, roton-like pair density waves are resolved both in the vortex and superconducting states of $CsV_3Sb_5$ by STM experiments (see Fig. 3c) [33]. This again points to an unconventional nature of the superconductivity. Nevertheless, a few studies favor a conventional superconducting picture. This is supported by the observations of reduced Knight shift below $T_c$, and a typical Hebel–Slichter coherence peak just below $T_c$ in nuclear magnetic resonance (NMR) and nuclear quadrupole resonance (NQR) experiments of $CsV_3Sb_5$ [75], as presented in Fig. 3e. Additionally, low-temperature STM experiments down to an effective electron temperature of 170 mK have revealed multiple superconducting gaps including U- and V-shaped gaps, in accordance with the multiband electronic structure [76]. Importantly, nonmagnetic impurities cannot induce in-gap states, indicating a sign-preserving or $s$-wave character of the superconducting state (see Fig. 3d) [76].

Besides the above-mentioned complex features of the charge order and superconductivity, the mutual interactions between CDW and superconductivity are also rather intricate. The coexistence of CDW and superconductivity may point to a common origin of these two instabilities. On the other hand, competitions also arise naturally since the two orders struggle for the same density of states at the Fermi level. As shown in Fig. 3f, for $CsV_3Sb_5$ under moderate pressures, $T_{CDW}$ drops rapidly while $T_c$ rises quickly, suggesting competing interactions between CDW and superconductivity [34–36,39]. By further increasing pressure, the CDW phase vanishes continuously and becomes invisible above 2 Gpa. On the other hand, $T_c$ is reduced slightly above ~ 0.7 Gpa followed by rapid enhancements above ~ 1 Gpa, giving rise to a striking double-dome structure. Similar features are also seen in $KV_3Sb_5$ and $RbV_3Sb_5$ [37,40,77]. A switching from low-pressure nodal to high-pressure nodeless superconductivity is also suggested in $RbV_3Sb_5$ [78]. Moreover, above ~ 14 Gpa, another dome-like superconducting phase emerges in $CsV_3Sb_5$ and survives up to 150 Gpa [35]. The nature of the pressure-induced reemergence of the second superconducting dome is still elusive. Possibilities including pressure-induced structural phase transition, two-dimensional to three-dimensional crossover have been suggested by X-ray diffraction measurements [35,79,80]. Enhanced electron-phonon coupling is also suggested in the second dome [35,79,80]. In addition, chemical doping can also effectively tune CDW and superconductivity. The CDW is suppressed by doping and superconducting double-dome is also seen in $CsV_{3-x}Ti_xSb_5$ and $CsV_3Sb_{5-x}Sn_x$ [81,82]. But only a single superconducting phase is observed in $Cs(V_{1-x}Nb_x)_3Sb_5$ [83] and another study on $CsV_{3-x}Ti_xSb_5$ [84]. The peculiar double-dome superconducting phase diagrams obtained by pressure tuning and doping in $A$V$_3$Sb$_5$ is reminiscent to those of high-temperature cuprate superconductors, offering another prominent channel to study the interplay of intertwined charge and superconducting orders.

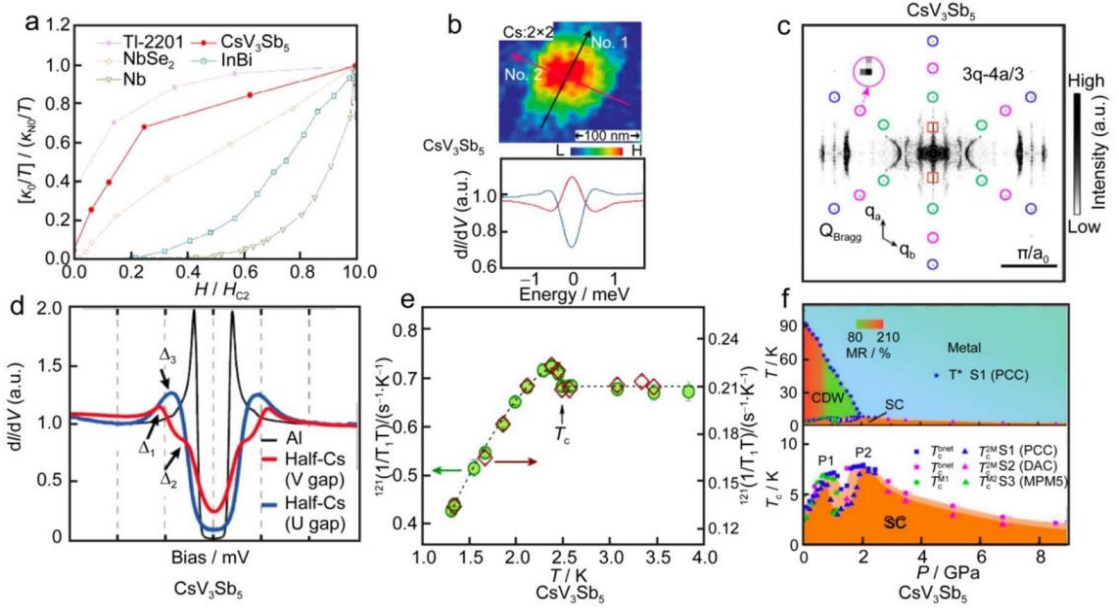

**Fig. 3** Superconducting properties of $A$V$_3$Sb$_5$. **a** Magnetic field dependence of the residual thermal conductivity, which suggests nodal superconductivity in CsV$_3$Sb$_5$. **b** Zero energy modes seen in the vortex core of CsV$_3$Sb$_5$ by STM. **c** Identification of pair density wave at the $\mathbf{Q}_{3q\text{-}4a/3}$ vector of CsV$_3$Sb$_5$ using STM. **d** Multigap structures showing both U-shape and V-shape gaps captured by STM in CsV$_3$Sb$_5$. **e** Observation of the Hebel–Slichter coherence peak in NQR experiments, which suggest an $s$-wave superconducting nature in CsV$_3$Sb$_5$. **f** Temperature-pressure phase diagrams of CsV$_3$Sb$_5$. A two-dome shaped superconducting phase is found. **a** Data from Ref. [29]. **b** Data from Ref. [30]. **c** Reproduced with permission from Ref. [33]. Copyright 2021 Springer Nature. **d** Adapted with permission from Ref. [76]. Copyright 2021 APS. **e** Data from Ref. [75]. Copyright 2021 IOP Publishing. **f** Data from Ref. [34].

# 5 Electrical transport properties

## 5.1 Quantum oscillations

When metals are subject to strong magnetic fields and low temperatures, various physical properties oscillate periodically as a function of magnetic field $B$ with a periodicity of $1/B$ due to the quantization of energy into Landau levels [85]. These oscillating properties include magnetization (de Haas-van Alphen effect), magnetoresistance (Shubnikov-de Haas oscillations), magnetostriction, magneto-Seebeck, etc. From these quantum oscillations, one can extract information about the extremal area of Fermi surfaces, electron effective mass, and topological properties of the band structure. Fig. **4** shows the Shubnikov-de Haas (SdH) oscillations in magnetoresistance (MR) observed for $AV_3Sb_5$. Clear SdH oscillations are seen in all three members. The oscillating components are better resolved after subtracting a polynomial background for MR measured at each temperature. The oscillation frequencies can be obtained accordingly by fast Fourier transform (FFT) analysis of the oscillatory parts (see Figs. 4b, d and f). In $KV_3Sb_5$, as shown in Figs. 4a, b, sizable SdH oscillations appear in the CDW state below 20 K and above 4 T [25]. Two frequencies are clearly identified at $F1$=34.6 T and $F2$=148.9 T. From the obtained frequencies, one can estimate the extremal area $A_F$ of the Fermi pockets via the Onsager relation:

$$F = (\hbar/2\pi e)A_F, \qquad (1)$$

where $F$ is the quantum oscillation frequency, $\hbar$ is the reduced Planck constant, $e$ is the electronic charge. The estimated area only accounts for a very small portion (0.2% and 0.9% for $F1$ and $F2$, respectively) of the in-plane first Brillouin zone. The in-plane lattice parameter $a$=0.54818 nm of $KV_3Sb_5$ has been used in the calculation [20]. The extremely small extremal area of the Fermi surfaces suggests that the transport properties are dominated by small pockets near the zone boundaries. This is further evidenced by the small effective electron mass ($m^*$) obtained by analyzing the temperature-dependent intensities of $F1$ and $F2$ using the Lifshitz-Kosevich approximation,

$$I(T) \approx X/(\sinh X), \qquad (2)$$

where $X = 14.69m^*T/B$, $B$ is normally taken as the average magnetic field of the field range used in FFT analysis. As shown in the inset of Fig. 4b, a small effective mass of $m^* = 0.125\ m_e$ ($m_e$ is the free electron mass) is obtained for $F1$. Similarly, two frequencies of $F_\alpha$=33.5 T and $F_\beta$=117.2 T are found in $RbV_3Sb_5$, as shown in Figs. 4c, d [22]. And a small effective electron mass of $m^* =$

$0.091\ m_e$ is also obtained for the $F_\alpha$ band. For CsV$_3$Sb$_5$, four frequencies locating at 18, 26, 72 and 92 T have been identified (see Figs. 4e, f) [26]. The effective masses estimated for the first two frequencies read as $m^* = 0.028\ m_e$ and $0.031\ m_e$, respectively. Similar results are found in a high-resolution MR study, which also resolves high frequency orbits up to ~2000 T in CsV$_3$Sb$_5$ [44]. The high-frequency parts are likely linked to the electron pocket located at the zone center (Γ point) [44]. Still, the dominant features are small extremal area and light effective electron mass, which point to dominant roles played by the small pockets and Dirac-like bands invoking the V-orbitals near the $M$ point [25,44].

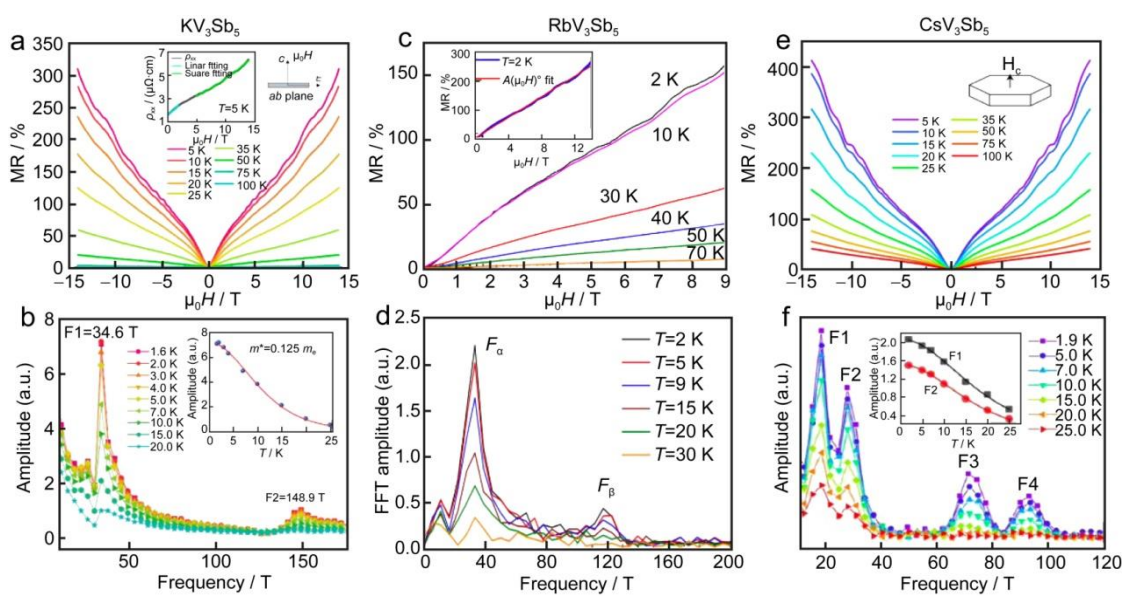

**Fig. 4** Quantum oscillations of $A$V$_3$Sb$_5$. **a** Magnetoresistance of KV$_3$Sb$_5$, which shows clear quantum oscillations at low temperatures. **b** FFT analysis of the quantum oscillation data shown in **a**. **c-f** Quantum oscillations observed for RbV$_3$Sb$_5$ and CsV$_3$Sb$_5$. **a** and **b** Data from Ref. [25]. **c** and **d** Data reproduced with permission from Ref. [22]. Copyright 2021 IOP Publishing. **e** and **f** Reproduced with permission from Ref. [26]. Copyright 2021 APS.

## 5.2 Anomalous Hall effect

One of the most intriguing properties of $A$V$_3$Sb$_5$ is the emergence of giant AHE in the charge ordered state without invoking any long-range magnetic orders. The absence of long-range magnetism has been verified by $\mu$SR experiments [86]. As displayed in Fig. 5a, b, large AHE is seen both in KV$_3$Sb$_5$ and CsV$_3$Sb$_5$ [25,26]. Note that, to better resolve the AHE, local backgrounds have been subtracted from the raw data. But no clear signature of AHE has been reported in RbV$_3$Sb$_5$, to our knowledge. At low temperatures, both the AHE conductivity ($\sigma_{AHE}$) and the anomalous Hall ratio ($\sigma_{AHE}/\sigma_{xx}$) of KV$_3$Sb$_5$ and CsV$_3$Sb$_5$ reach large values of ~ $10^4$ $\Omega^{-1}$ cm$^{-1}$ and ~ 2%, which are one order of magnitude larger than those found in typical ferromagnetic metals, such as Fe and Ni [25,26]. Originally, the observed large AHE in KV$_3$Sb$_5$ was attributed to extrinsic mechanisms, such as enhanced skew scattering arising from Dirac-like quasiparticles of the kagome lattice [25]. In CsV$_3$Sb$_5$, it is suggested that both extrinsic skew scattering and intrinsic large Berry curvature contribute to AHE [26]. Importantly, the AHE appears concomitantly with the CDW instability (see Fig. 5c), suggesting intimate coupling between AHE and CDW [26]. Considering the unconventional nature of CDW, as we have discussed in Section 3, the AHE may also arise from the intrinsic time-reversal breaking nature of the charge order.

The AHE can be easily tailored via gating and chemical doping, as displayed in Fig. 5d, f [69,83]. Using a protonic gate method, the carrier density ($n$) of CsV$_3$Sb$_5$ thin flakes can be largely modulated, ranging from electron dominated to hole dominated regions (see Fig. 5d) [69]. The AHE is seen both in the electron and hole doped sides. Away from these two regions, AHE diminishes, suggesting that the AHE is extremely sensitive to carrier concentration and the position of Fermi level [69]. The large AHE reaching ~$10^4$ $\Omega^{-1}$ · cm$^{-1}$ is only found in a narrow hole doped region around $n \sim 2.5 \times 10^{23}$ cm$^{-3}$, when the Fermi level is sitting inside the CDW gap. In the hole doped side, the large AHE is attributed to extrinsic skew scattering stemming from magnetic fluctuations [69]. In the electron doped region, the AHE likely originates from intrinsic Berry curvature of the band topology. In bulk CsV$_3$Sb$_5$ samples, the AHE gradually weakens upon doping by Nb, as seen in Fig. 5f [83]. The AHE becomes nearly invisible in Cs(V$_{1-x}$Nb$_x$)$_3$Sb$_5$ for $x$=0.07 even at a low temperature of 5 K. Notably, the CDW transition temperature $T_{CDW}$ is also suppressed continuously with increasing Nb doping. This again implies that the AHE is closely related to CDW. It turns out that the electron pocket centered at the $\Gamma$ point is enlarged by doping. And the van Hove singularity siting below the *M* point in undoped sample shifts upwards and passes the Fermi energy with increasing doping [83]. These results imply that the AHE is linked to the Fermi level and the pockets around the *M* point.

Since the AHE is sensitive to carrier concentration and Fermi energy, the AHE can also become vanishingly small in as grown *A*V$_3$Sb$_5$ samples due to the existence of vacancies at *A* sites and paramagnetic impurities [20,26]. As shown in Fig. 5f, the AHE indeed is negligible in CsV$_3$Sb$_5$ samples with relatively low residual resistivity ratio (RRR) [87]. In this case, the raw data of Hall resistivity $\rho_{yx}$ is dominated by multiband transport well inside the CDW phase. At high-temperature above 50 K, $\rho_{yx}$ scales linearly with magnetic field $B$ with a negative slope, indicating dominant electron-like transport. Below 50 K, nonlinear field dependence of $\rho_{yx}(B)$ starts to develop, suggesting involvement of multiband transport. Indeed, the sublinear $\rho_{yx}(B)$ curves measured below 50 K can be well described by a simple two-band fitting (see solid lines in Fig. 5f) [87]. Notably, the slope of $\rho_{yx}(B)$, i.e., the Hall coefficient, switches to positive values below a characteristic temperature $T^* \sim 35$ K (see Fig. 5f). This indicate that, hole-like carriers become dominant in the Hall resistivity below $T^*$. The competitions between electron-like and hole-like bands also produce sizable Nernst signal, which we will discuss in Section 6.3. Note that this sign switching in the temperature dependence of the Hall coefficient is generic in $A$V$_3$Sb$_5$, regardless of the appearance of AHE and sample quality. Though slightly different values of $T^*$ are found in KV$_3$Sb$_5$ and RbV$_3$Sb$_5$, which are $T^* \sim 25$ and 45 K, respectively [22,25,88]. Therefore, although the AHE is very sensitive to carrier concentration, Fermi energy and sample quality, etc., the multiband transport is, however, robust and appears to be generic in all three members [88]. This is certainly resulted from the multiband nature of electronic structure as we have discussed in Section 2. The origin of such a sign change in the temperature-dependent Hall coefficient is still unknown. It may originate from temperature induced Lifshitz transition [69], enhancement of hole mobility below $T^*$ [22,26], or even correlate with additional symmetry breaking below $T^*$, such as electronic nematicity (see Section 5.3).

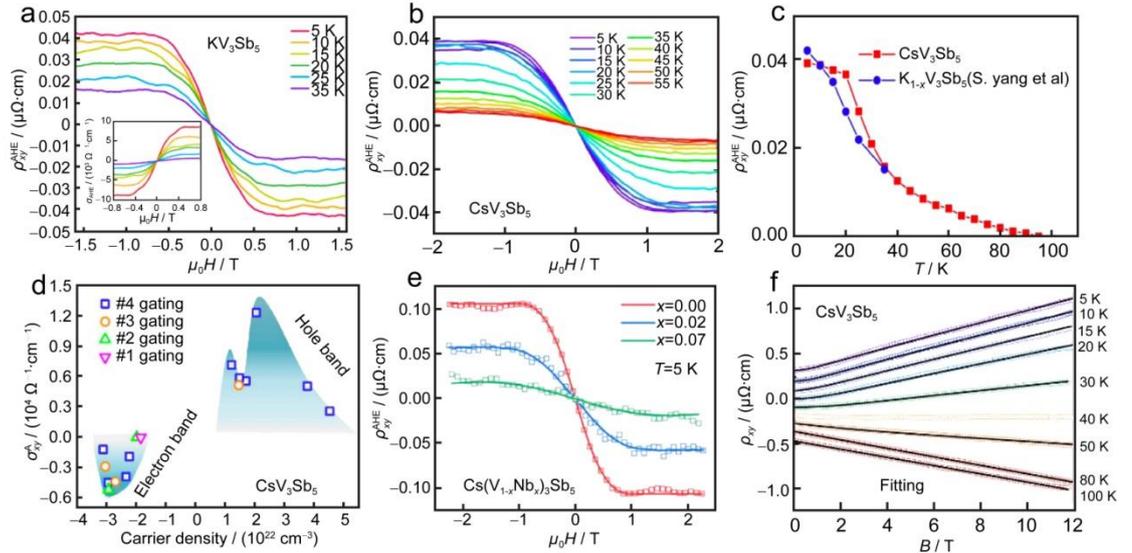

**Fig. 5** Anomalous Hall effect of $A$V$_3$Sb$_5$. **a** Anomalous Hall resistivity and conductivity (inset) found in the CDW phase of KV$_3$Sb$_5$. **b** Anomalous Hall resistivity of CsV$_3$Sb$_5$. **c** Temperature dependence of the anomalous Hall resistivity. The AHE occurs concomitantly with CDW in CsV$_3$Sb$_5$. **d** Gate tuned AHE in thin flakes of CsV$_3$Sb$_5$. The AHE depends strongly on carries density. **e** The AHE disappears gradually upon Nb doping in CsV$_3$Sb$_5$. **f** Multiband effects in CsV$_3$Sb$_5$. The AHE becomes less evident in samples with relatively low RRR values. A simple two-band fit (solid lines) can describe the Hall resistivity data well. **a** Data from Ref. [25]. **b** and **c** Adapted with permission from Ref. [26]. Copyright 2021 APS. **d** Data from Ref. [69]. **e** Reproduced with permission from Ref. [83]. Copyright 2022 APS. **f** Adapted with permission from Ref. [87]. Copyright 2021 APS.

## 5.3 Electronic nematicity

In addition to CDW and superconductivity, there may exist additional symmetry breaking both in the CDW and superconducting phases. As shown in Fig. 6a-c, the *c*-axis resistivity ($\rho_c$) of CsV$_3$Sb$_5$ shows a clear twofold rotational symmetry both in the charge ordered state and superconducting vortex state [31]. This indicates that the rotational symmetry is reduced from $C_6$ to $C_2$ in these phases, suggesting the emergence of electronic nematicity that breaks rotational symmetry while preserves translational symmetry. The emergence of nematicity in the superconducting state may also suggest an unconventional nematic superconducting nature. The $C_2$ symmetric *c*-axis resistivity starts to develop below $T_{CDW}$, and picks up strength rapidly below ~ 40 K. The appearance of nematicity is further evidenced by STM, NMR and elastoresistivity experiments, which suggest that the long-range nematic order is established below $T_{nem}$ ~ 35 K [32]. As presented in Fig. 6d, the ($m_{11} - m_{12}$) component of the elastoresistivity tensor clearly peaks at $T_{nem}$, which is a signature of nematic instability. Here, the ($m_{11} - m_{12}$) component represents the nematic susceptibility in the $E_{2g}$ channel of the $D_{6h}$ point group. The nematic susceptibility increases gradually just below $T_{CDW}$, indicating close connections between CDW and nematic fluctuations. The three-dimensional charge order can induce a symmetry reduction from $C_6$ to $C_2$ by stacking the distorted kagome layers either in the form of SD/SD, ISD/ISD or SD/ISD (see Section 3) [30–32,47,55,56,66,67]. Such $C_2$ distortions can easily couple to the nematic order and enhance nematic fluctuations. Establishment of long-range nematic order below $T_{nem}$ may also leave fingerprints in other properties. The muon spin relaxation rates show additional enhancements below ~ 30 K [27]. In Raman scattering experiments, two additional modes are seen below ~ 30 K [43]. Interestingly, the nematic transition occurs at a similar temperature compared to the sign reversal temperature $T^*$ of the Hall resistivity. The correlation between these two effects remains to be elucidated in further studies.

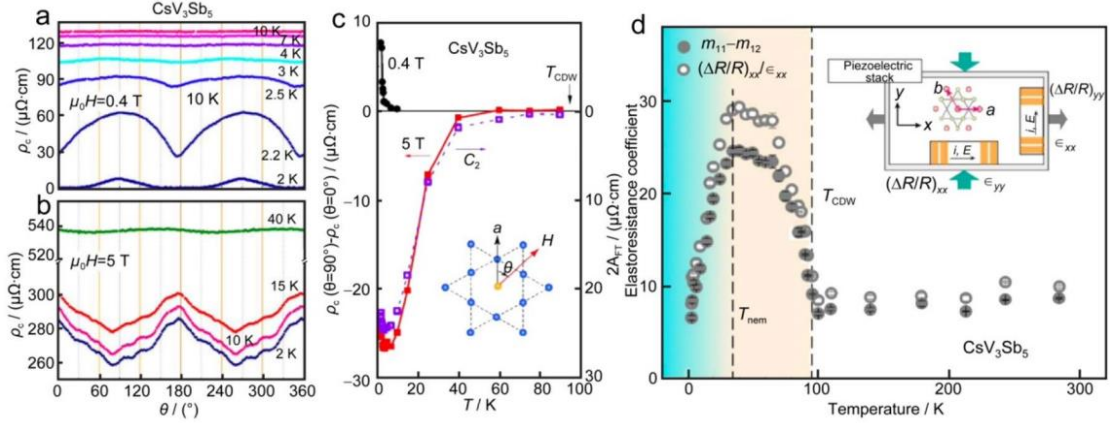

**Fig. 6** Electronic nematicity in $A$V$_3$Sb$_5$. **a** and **b** The $c$-axis resistivity of CsV$_3$Sb$_5$ shows a $C_2$ symmetric behavior when a magnetic field is rotating within the $ab$-plane, both in the CDW and superconducting states. **c** Temperature dependence of resistivity anisotropy, which survives up to $T_{CDW}$. **d** Temperature dependence of the elastoresistivity coefficient found in CsV$_3$Sb$_5$. The nematic susceptibility in the $E_{2g}$ ($m_{11} - m_{12}$) channel develops below $T_{CDW}$ and peaks at $T_{nem}$, suggesting emergence of a long-range nematic order below $T_{nem}$. **a, b** and **c** Data from Ref. [31]. **d** Reproduced with permission from Ref. [32]. Copyright 2022 Springer Nature.

## 6 Thermoelectric properties

In the presence of a thermal gradient along the **x** direction ($-\nabla T \parallel$ **x**), charge carriers in metals or semiconductors diffuse from the hot end to the cold end, building up an electric field along the thermal current direction (**E** $\parallel$ **x**). This effect is called the longitudinal thermoelectric effect or Seebeck effect [89]. The Seebeck coefficient is defined as the ratio between the electric field and thermal gradient, $S_{xx} = E_x/\partial T_x$. When a magnetic field is applied along the **z** direction (**B** $\parallel$ **z**), a transverse electric field along the **y** direction (**E** $\parallel$ **y**) can be produced by a thermal gradient in the **x** direction, in analogous to the Hall effect. This effect is termed as the transverse thermoelectric effect or Nernst effect, which is defined as the ratio between the traverse electric field ($E_y$) and longitudinal thermal gradient ($\partial T_x$), $S_{yx} = E_y/\partial T_x$. Here, the notation of Cartesian coordinate has been used, i.e., **x**, **y**, **z** are orthogonal to each other. The Seebeck effect and Nernst effect probe the entropy flow of charge carriers, and measure the energy derivative of the electrical conductivity at the Fermi energy. The thermoelectric properties are thus extremely sensitive to nontrivial topological bands near the Femi level [90,91]. The thermoelectric probes have been extensively used to investigate unusual properties of quantum materials, such as high-temperature superconductors and topological materials [92–95]. In this section, we briefly summarize the thermoelectric properties of $A$V$_3$Sb$_5$.

## 6.1 Magneto-Seebeck effect

In Fig. 7, the Seebeck effect ($S_{xx}$) of CsV$_3$Sb$_5$, together with its response to external magnetic fields, are presented. Similar results have also been observed in KV$_3$Sb$_5$ and RbV$_3$Sb$_5$ [88]. As seen in Fig. 7a, the Seebeck signal of CsV$_3$Sb$_5$ shows a sudden jump around $T_{\text{CDW}}$ = 94 K [87]. Moreover, negative values of $S_{xx}$ are seen at most temperatures in zero field, except below about 7 K. This indicates dominant roles played by electron-like carriers in the thermoelectric properties. Notably, $S_{xx}$ becomes positive below 7 K and reaches zero in the superconducting state below $T_c$=2.7 K. Such a sign change in the temperature dependence of Seebeck signal is typically a result of multiband transport, in accordance with the multiband transport seen in Hall resistivity (see Fig. 5f). Note that the Hall coefficient changes sign at a different temperature $T^*$ ~ 35 K in CsV$_3$Sb$_5$. This difference is not unexpected since electron-like and hole-like carriers compete in different fashions in electrical and thermoelectric channels. Generally, the total Seebeck signal of a multiband metal is the weighted contribution from each band,

$$S_{\text{tot}} = \sum \sigma_i S_i / \sum \sigma_i, \tag{3}$$

with $\sigma_i = n_i e \mu_i$, $S_i$, $n_i$ and $\mu_i$ being the electrical conductivity, Seebeck coefficient, carrier density and mobility of the $i^{\text{th}}$ band [89]. The contribution from each band can be approximated as $S_i = \pm N_i(E_F)\pi^2 k_B^2 T/3n_i e$ via the Mott relation using the free electron gas and energy-independent relaxation approximations [96]. Here, $N_i(E_F)$ is the density of states (DOS) at the Fermi level, $k_B$ is the Boltzmann constant. The positive (negative) sign is contributed from hole (electron) bands. Therefore, the total Seebeck signal depends on DOS, carrier density and mobility of each band. Below

$T^*$, a two-band fitting of the Hall resistivity shows that the hole concentrations become dominant, leading to a positive sign in Hall coefficient [87]. The mobility of the electron band, on the other hand, is much higher than that of the hole band, giving rise to negative values of Seebeck signal even below $T^*$. The sudden enchantments of $S_{xx}$ just below $T_{CDW}$ may also suggest that the DOS of the hole band is gapped more compared with that in the electron pocket [88]. The sign switching temperature in $S_{xx}(T)$ shifts to higher temperatures in the presence of magnetic fields, and moves to ~ 20 K in 13 T (see the inset in Fig. 7a). This likely suggests that the DOS of the hole band is enhanced by the application of magnetic fields.

The Seebeck coefficient of CsV$_3$Sb$_5$ shows substantial response to magnetic field only in the CDW state, as shown in Fig. 7a, b. The magneto-Seebeck signal $\Delta S_{xx} = S_{xx}(13\ \text{T}) - S_{xx}(0\ \text{T})$ only becomes visible below $T_{CDW}$ and increases gradually at lower temperatures. Similar temperature dependence of the magnetoresistance $\Delta\rho_{xx} = \rho_{xx}(13\ \text{T}) - \rho_{xx}(0\ \text{T})$ is also found (see Fig. 7b). These effects are likely rooted from the emergent time-reversal symmetry breaking of the charge order. Moreover, as presented in Fig. 7c,e, pronounced quantum oscillations are seen in the magneto-Seebeck effect due to its high sensitivity to DOS near the Fermi level [87,97]. As shown in Fig. 7d, f, four frequencies locating around 18, 28, 72 and 88 T can be identified, and the estimated effective electron masses of these bands are very small (see the inset in Fig. 7d), agreeing well with those found in SdH oscillations [26,87,88,97]. The light effective masses likely point to the linearly dispersive Dirac-like bands near the zone boundaries, as we have also discussed earlier. Therefore, the thermoelectric probe is another prominent channel to uncover the fermiology and topological features of the underlying electronic band structure.

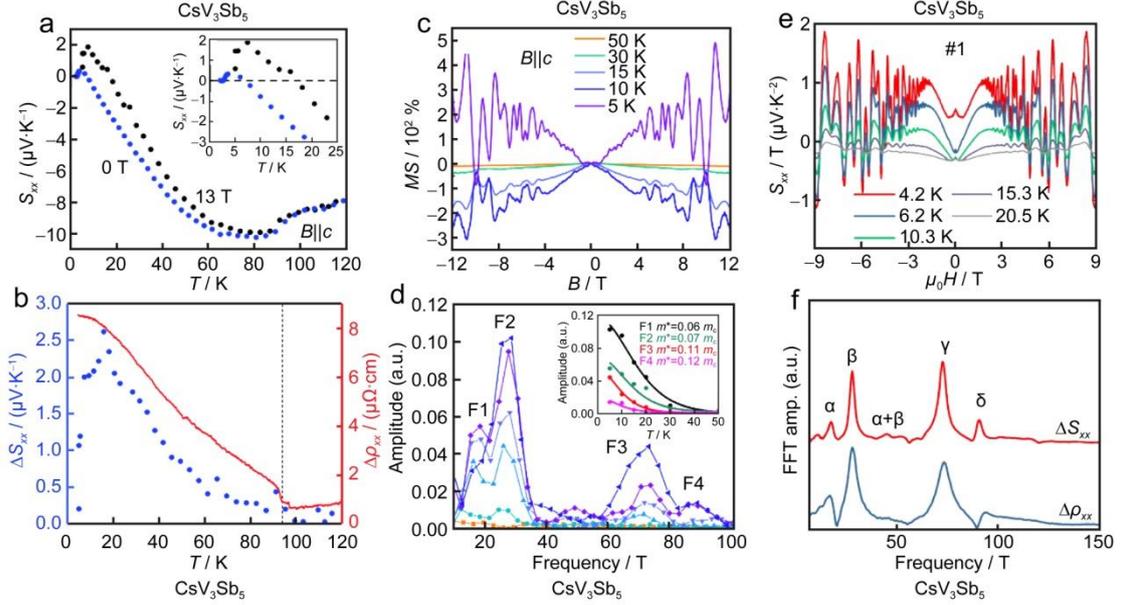

**Fig. 7** Magneto-Seebeck effect of $CsV_3Sb_5$. **a** Temperature dependence of the Seebeck signal of $CsV_3Sb_5$ measured in zero field and 13 T with magnetic fields applied out-of-plane. The inset shows that the Seebeck data crosses zero at ~ 7 and 20 K in 0 and 13 T, respectively. **b** Temperature-dependent magneto-Seebeck $\Delta S_{xx} = S_{xx}(13\ \text{T}) - S_{xx}(0\ \text{T})$ and magnetoresistance $\Delta \rho_{xx} = \rho_{xx}(13\ \text{T}) - \rho_{xx}(0\ \text{T})$. **c** and **e** Quantum oscillations observed in the magneto-Seebeck effect. **d** and **f** FFT spectra evaluated from the quantum oscillation data shown in **c** and **e**. **a-d** Adapted with permission from Ref. [87]. Copyright 2021 APS. **e** and **f** Data reproduced with permission from Ref. [97]. Copyright 2022 APS.

## 6.2 Anomalous Nernst and Ambipolar Nernst effects

Although the electrical Hall effect and thermoelectric Nernst effect share similarities, they differ quite significantly. For example, in systems carrying nontrivial band topology with sizable Berry curvature ($\Omega_B$), AHE measures the average Berry curvature across the entire occupied states, whereas anomalous Nernst effect (ANE) only probes the Berry curvature in the vicinity of the Fermi energy. This is readily seen by noticing that the anomalous Hall conductivity ($\sigma_{xy}^{AHE}$) and the anomalous Nernst conductivity ($\alpha_{xy}^{ANE}$) are connected to $\Omega_B$ in the forms of [90,98,99],

$$\sigma_{xy}^{AHE} = \frac{e^2}{\hbar} \int \frac{d^3k}{(2\pi)^3} f(k) \Omega_B(k) \approx \frac{e^2}{\hbar} \langle \frac{\Omega_B}{\lambda_F^2} \rangle, \qquad (4)$$

$$\alpha_{xy}^{ANE} = \frac{ek_B}{\hbar} \int \frac{d^3k}{(2\pi)^3} s(k)\Omega_B(k) \approx \frac{ek_B}{\hbar} \langle \frac{\Omega_B}{\Lambda^2} \rangle. \tag{5}$$

Here, $f(k)$ represents the Fermi-Dirac distribution, $s(k)$ is the entropy density which is nonzero for electronic states near the Fermi level only, $\lambda_F$ is the Fermi wavelength and $\Lambda = (h^2/2\pi m k_B T)$ stands for the de Broglie thermal wavelength, $h$ is the Planck constant, $m$ is the electron mass. Thus, all occupied sates contribute to $\sigma_{xy}^{AHE}$. On the other hand, $\alpha_{xy}^{ANE}$ is only sensitive to states living close to the Fermi energy within a thickness of $\Lambda$, since these states carry finite entropy. The ANE can thus serve as a sensitive probe in detecting topological states, as found in various topological materials [93]. As discussed in Section 5.2, giant AHE emerges in the CDW phase of $AV_3Sb_5$. It is also of great interest to explore the thermoelectric counterpart, i.e., ANE.

Fig. 8 shows representative Nernst results obtained in $CsV_3Sb_5$ [87,97,100]. As shown in Fig. 8a, 'S'-shaped field dependence of the Nernst signal is observed at low temperatures well inside the CDW state, indicating the appearance of ANE in $CsV_3Sb_5$ [97]. Clear quantum oscillations are also seen above 3 T. The observation of sizable ANE implies considerable contributions from Berry curvature, likely originated from the Dirac points near the Fermi level [97]. Note that clear signatures of ANE are found in high-quality samples possessing low degrees of defects as evidenced by extremely large values of residual resistivity ratio RRR = $\rho_{xx}$(300 K)/$\rho_{xx}$(5 K) ~ 325 [97]. The carrier mobility is found to be as high as $\mu$ ~ $10^5$ cm$^2$ V$^{-1}$ · s$^{-1}$. Thus, the ANE is also sensitive to sample quality, like the AHE. For samples with lower mobilities, the amplitudes of ANE are significantly reduced [97]. This suggests that extrinsic mechanisms, such as skew scattering, are also contributing to ANE. Unlike the AHE which appears concurrently with the charge order, the ANE only becomes visible below ~ 30 K, as seen more clearly in Fig. 8b. Intriguingly, as we have discussed before, the nematic transition and the sign reversal in the temperature-dependent Hall coefficient occur at similar temperatures. The appearance of sizable ANE below ~ 30 K likely has the same origin, i.e., additional symmetry breaking well inside the CDW state [97].

Another study shows similar results, as shown in Fig. 8c [100]. Again, sizable antisymmetric 'S'-shaped Nernst effect appears below ~ 30 K. Interestingly, the temperature dependence of the total Nernst signal peaks in a temperature interval of 30 ~ 40 K, depending on the strength of magnetic field (see Fig. 8d). In high-magnetic field above 3 T, the Nernst effect is dominated by ordinary contributions. This peculiar temperature dependence of ordinary Nernst effect suggests a multiband transport nature, since the Nernst signal is expected to be vanishing in a single band metal due to Sondheimer cancellation [95,101]. The multiband effects are seen more clearly in samples having less contributions from ANE, as presented in Fig. 8e, f [87]. The field dependence of Nernst signal is basically linear above 50 K. Sublinear effects appear below 40 K and weak 'S'-shaped signal only becomes visible below 10 K. The largely reduced contributions from ANE are likely due to the relatively low residual resistivity ratio of ~ 52 for the samples studied in Ref. [87]. Similarly, the AHE is also less evident compared to that in samples with higher values of RRR. Apparently, both AHE and ANE are sensitive to impurities, carrier concentrations and Fermi energy.

Following Wang *et al.*, one can express the Nernst signal as [101],

$$S_{yx} = S_{xx}\left(\frac{\alpha_{yx}}{\alpha_{xx}} - \frac{\sigma_{yx}}{\sigma_{xx}}\right) = S_{xx}(\tan\theta_\alpha - \tan\theta_H), \tag{6}$$

where $\alpha_{ij}$ and $\sigma_{ij}$ ($i, j = x, y$) are thermoelectric and electrical conductivity tensors, $\theta_\alpha = \alpha_{yx}/\alpha_{xx}$ and $\theta_H = \sigma_{yx}/\sigma_{xx}$ are Hall-like angles. In one band metals assuming energy-independent conductivity, $\theta_\alpha$ and $\theta_H$ typically cancels out (Sondheimer cancellation), giving negligible Nernst signal. Things become different in a multiband system and the Nernst signal of a simple two-band metal now reads as,

$$S_{yx} = S_{xx}\left(\frac{\alpha_{yx}^h + \alpha_{yx}^e}{\alpha_{xx}^h + \alpha_{xx}^e} - \frac{\sigma_{yx}^h + \sigma_{yx}^e}{\sigma_{xx}^h + \sigma_{xx}^e}\right), \tag{7}$$

where $\alpha_{ij}^{h(e)}$ and $\sigma_{ij}^{h(e)}$ are conductivity contributions from hole (electron) bands. For compensated bands, $\sigma_{yx}^h = -\sigma_{yx}^e$, the Hall effect vanishes. On the other hand, the Nernst effect is enhanced in this case compared to that in a single band system, since $\alpha_{yx}^h$ and $\alpha_{yx}^e$ share the same sign, and the second term in Eq. (7) goes to zero. The enhanced Nernst effect due to compensation of electron-like and hole-like charge carriers is called ambipolar Nernst effect, which is typically found in compensated semiconductors [102], and has also been reported in a well-known CDW superconductor 2*H*-NbSe$_2$ [103].

As shown in Fig. 8f, the ambipolar Nernst effect is also realized in CsV$_3$Sb$_5$ samples when multiband effects dominate. It is seen that in the temperature dependence of the Nernst signal, a broad peak appears around $T^*$ ~ 35 K. Notably, the Hall coefficient also changes sign near $T^*$. Therefore, when compensation between electron-like and hole-like bands happens, the Nernst signal reaches maximum, indicating an ambipolar transport nature. The experimental data can be well reproduced by Eq. (6) (see the blue solid line in Fig. 8f), suggesting that the Sondheimer cancellation is indeed avoided by ambipolar flow of charge carriers with different signs. The ambipolar Nernst effect is also seen in KV$_3$Sb$_5$ and RbV$_3$Sb$_5$, pointing to a generic feature shared among the $A$V$_3$Sb$_5$ members [88]. Therefore, the kagome $A$V$_3$Sb$_5$ family provides another prominent metallic system to study the interplay of CDW, superconductivity and ambipolar transport effects, in addition to the famous CDW superconducting system 2$H$-NbSe$_2$.

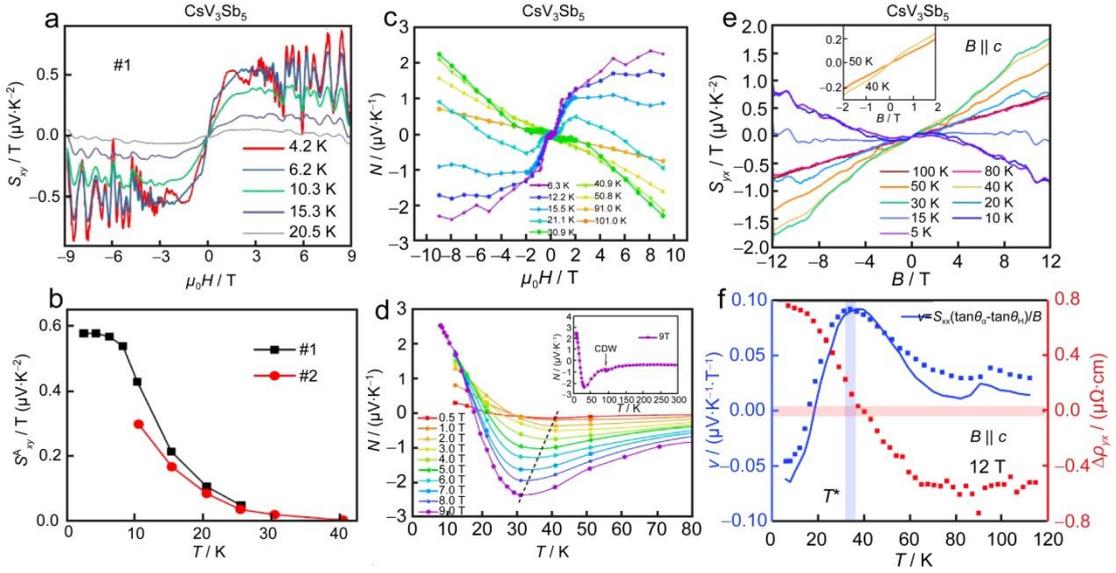

**Fig. 8** Anomalous Nernst and ambipolar Nernst effects of CsV$_3$Sb$_5$. **a** and **c** 'S'-shaped ANE found at low temperatures. **b** Temperature dependence of the ANE, which becomes sizable below about 30 K. **d** The total Nernst signal plotted as a function of temperature. The amplitude of Nernst data peaks around 30-40 K depending on magnetic fields. **e** Multiband dominated Nernst effect found in samples with relatively low RRR. **f** Temperature dependence of the Nernst coefficient in comparison with the Hall resistivity. The Nernst coefficient reaches maximum at $T^* \sim 35$ K, at which temperature the Hall resistivity changes sign, suggesting the appearance of ambipolar Nernst effect. **a** and **b** Adapted with permission from Ref. [97]. Copyright 2022 APS. **c** and **d** Reproduced with permission from Ref. [100]. Copyright 2022 APS. **e** and **f** Adapted with permission from Ref. [87]. Copyright 2021 APS.

## 7 Thermal transport properties

Both bosonic and fermionic excitations participate in heat conduction of solids. Thermal transport measurements can thus extract various information of quantum materials. For example, the longitudinal thermal conductivity ($\kappa_{xx}$) in the superconducting state can reflects the symmetry of superconducting order parameter [104]. As we have seen in Fig. 3a, the thermal conductivity measurements in the superconducting state have suggested the possible existence of nodal superconductivity [29]. The transverse thermal conductivity ($\kappa_{xy}$), i.e., the thermal Hall effect, has been extensively applied to study nontrivial low-energy excitations in quantum materials. Giant thermal Hall effects have been observed in high-temperature superconducting cuprates, multiferroics and various quantum magnets [105–109]. Half-integer quantized thermal Hall effect, a hallmark of quantum spin liquids, has been reported in a Kitaev quantum spin liquid candidate material $\alpha$-RuCl$_3$ [110–112]. The origins of these peculiar thermal Hall effects are still elusive. It is worth noting that phonons can also play important roles in the thermal Hall effect [113]. In this section, both the longitudinal and transverse thermal conductivities of CsV$_3$Sb$_5$ are briefly summarized.

Fig. **9**a presents the longitudinal thermal conductivity of CsV$_3$Sb$_5$ measured in the normal state above the superconducting transition [100]. It is seen that the electronic contributions evaluated via the Wiedemann-Franz $k_{el} = L_0 \sigma_{xx} T$ ($L_0$: the Lorenz number) only account partially for the total thermal conductivity. This indicates that phonons also contribute significantly to the heat conduction. Interestingly, $\kappa_{xx}$ decreases monotonically upon cooling until $T_{CDW}$, deviating from the typical $1/T$ temperature dependence of Umklapp scattering dominated phonon transport. This is likely caused by additional scattering events provided by charge fluctuations of the CDW, as also seen in other CDW materials [114,115]. In the long-range charge ordered state, charge fluctuations are quenched, leading to the recovery of the typical thermal transport behavior. As shown in Fig. 9b, sizable thermal magnetoconductivity $\kappa_{xx}(B)$ is only observed in the CDW state, agreeing well with that found in magnetoresistance and magneto-Seebeck effect. Moreover, sizable thermal Hall effect is found in the CDW phase (see Fig. 9c). In particular, non-linear field dependence of $\kappa_{xy}(B)$ develops below 40 K, which are signatures of anomalous thermal Hall effect (ATHE). Note that two 'S'-like features appear in low magnetic fields below 2 T and high magnetic fields above 6 T. The high-field sublinear $\kappa_{xy}(B)$ is attributed to multiband effects, and the low-field feature is likely coming from ATHE [100]. An enlarged view of the ATHE parts after subtracting local backgrounds is shown in Fig. 9d. It is seen that the ATHE mainly develops below 40 K, which is consistent with the ANE (see Fig. 8b). The temperature dependence of the total thermal Hall conductivity $\kappa_{xy}$ agrees quantitatively with the electronic contribution obtained from the Wiedemann-Franz $k_{xy}^{el} = L_0 \sigma_{xy} T$, as seen in Fig. 9e. In addition, the anomalous thermal Hall ratio $\kappa_{xy}^{ATHE}/\kappa_{xx}$ is comparable to the anomalous Hall ratio $\sigma_{xy}^{AHE}/\sigma_{xx}$ (see Fig. 9f). Thus, the thermal Hall effect is dominated by electronic part, and is intimately correlated with AHE, ANE, which are likely to originate from time-reversal symmetry breaking and nontrivial topological features.

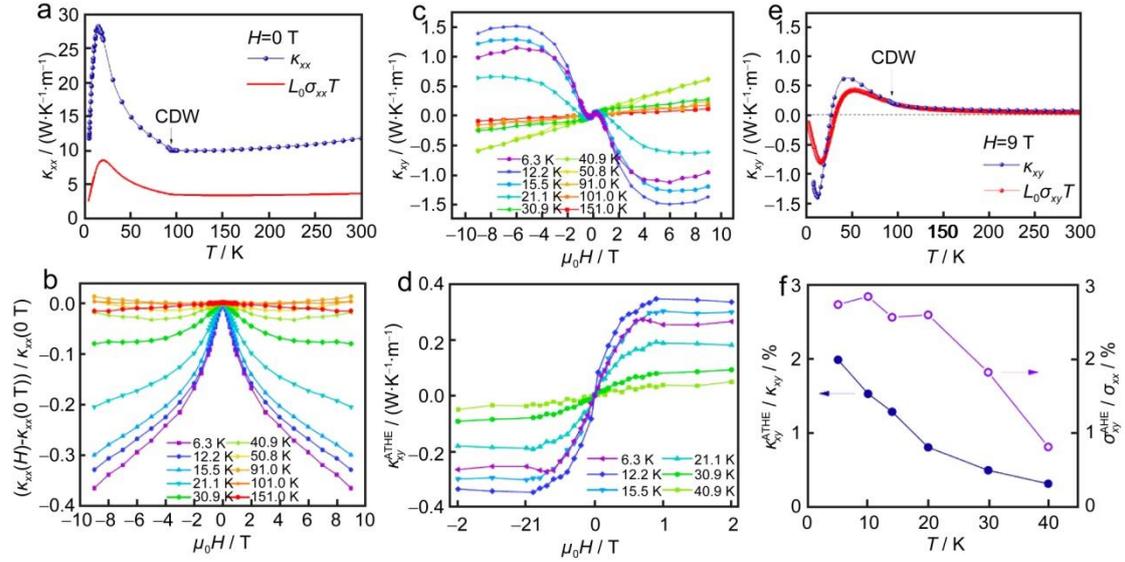

**Fig. 9** Thermal transport properties of CsV$_3$Sb$_5$. **a** The longitudinal thermal conductivity of CsV$_3$Sb$_5$ measured as a function of temperature. The electronic contributions are also shown for comparison (red solid line). **b** The thermal magnetoconductivity recorded at fix temperatures. **c** Thermal Hall conductivity measured at various temperatures. 'S'-shaped thermal Hall data appears below 40 K, indicating the existence of anomalous thermal Hall contributions. **d** The anomalous thermal Hall conductivity obtained from the data shown in **c** after subtracting local backgrounds. **e** Temperature dependence of the thermal Hall conductivity in comparison with the electronic contributions. **f** Comparison of anomalous thermal Hall ratio and anomalous Hall ratio. Reproduced with permission from Ref. [100]. Copyright 2022 APS.

# 7 Outlook

By summarizing the recent progress of the newly discovered kagome metals $A$V$_3$Sb$_5$, it becomes clear that this family is a prominent platform to study the intricate interplay of lattice geometry, electronic correlations and band topology. A variety of unusual physical phenomena arise in the $A$V$_3$Sb$_5$ compounds due to the emergence of intertwined unconventional charge, nematic and superconducting orders. In spite of extensive research effort, a few open questions still need to be addressed by further investigations.

The in-plane charge modulation of the CDW has been consistently determined to be $2 \times 2$. However, the exact form, including $2 \times 2 \times 2$, $2 \times 2 \times 4$, or coexistence of the two, of three-dimensional stacking remains to be resolved. This challenge is commonly faced by layered materials, in which different stacking sequences share similar energies. In real materials, different stacking forms often coexist, leading to the diverse experimental results detected by different techniques, as found here for $A$V$_3$Sb$_5$. Moreover, the driving force of the CDW instability is still elusive. Many experimental and theoretical works favor an electronic scenario, where Fermi surface nesting between the saddle points plays a dominant role. The importance of the electron-phonon coupling, however, cannot be ignored, as suggested by some other studies.

The nature of the superconducting state remains to be elucidated. Both conventional *s*-wave and unconventional nodal pictures have been suggested. Observations of pair density wave, Majorana bound states, and nematicity further complicates the problem. These seemly contradicting results may be reconciled by noting the possible time-reversal symmetry breaking effects found in the CDW state. The emergence of time-reversal symmetry breaking could allow gapless excitations even in a fully gapped superconducting state. The connection between superconductivity and CDW, and the nature of these two orders can be further unraveled with the assistance of external tuning parameters, such as uniaxial strain, hydrostatic pressure and chemical doping. It is also of great interests to uncover more nontrivial excitations arising from the interplay of superconductivity, time-reversal symmetry breaking and topological band structures.

Increasing experimental evidence has suggested a time-reversal symmetry breaking charge order. The theoretical proposed chiral flux phase can naturally explain the emergent time-reversal symmetry breaking effects. Still, direct confirmation of such a chiral flux phase is still lacking. In addition, CDW without chirality has also been experimentally reported. A unified picture reconciling these conflicting aspects is still challenging. It is also noteworthy that additional symmetry breaking effects, such as rotational symmetry breaking and electronic nematicity, likely occur well inside the CDW phase. The impacts of these symmetry breaking effects on CDW and superconductivity remain to be furtherly explored.

Concerning the unusual transport properties, the correlation between the observed giant AHE, time-reversal symmetry breaking, and nontrivial band topology is not clear. It appears that both AHE, ANE are sensitive to impurities, carrier concentration and the Fermi level, suggesting the important roles played by extrinsic mechanisms. The multiband transport and ambipolar Nernst effect, on the other hand, appear to be robust and generic in all three members of $A$V$_3$Sb$_5$. Moreover, in all three compounds, the Hall coefficient changes sign at a certain temperature $T^*$. The origin of such a sign reversal effect in the temperature-dependent Hall coefficient remains to be explored. Interestingly, in CsV$_3$Sb$_5$, long-range nematic order, sizable ANE and ATHE also emerge below ~ $T^*$. The impacts of the nematic transition on electronic band structure and transport properties, are waiting to be elucidated.

In summary, the discovery of the kagome metals $A$V$_3$Sb$_5$ has triggered extensive research activities both experimentally and theoretically. Motivated by the observations of various emergent quantum phenomena within one single system, novel topics and concepts have also been raised. Electrical and thermal transport studies have served as prominent channels to uncover the unusual properties of $A$V$_3$Sb$_5$. Combined with other techniques, we anticipate further rapid progress in unraveling the underlying intricate physics in $A$V$_3$Sb$_5$. Along with the exciting research activities in $A$V$_3$Sb$_5$, a few related compounds containing similar V-kagome lattice have been experimentally discovered, such as $R$V$_6$Sn$_6$ ($R$=Gd, Ho), $A$V$_6$Sb$_6$, $A$V$_8$Sb$_{12}$ and V$_3$Sb$_2$ [116–120]. With such diversity of research interests, a broader variety of intriguing quantum phases of matter are expected in these kagome materials, which are certainly helpful to resolve the interplay of geometric frustration, strong correlations and topological properties in kagome systems.

# Acknowledgement

We thank G. Wang and Y. Liu at the Analytical and Testing Center of Chongqing University for technical support. The authors acknowledge the support by National Natural Science Foundation of China (Grant No. 11904040), Chongqing Research Program of Basic Research and Frontier Technology, China (Grant No. cstc2020jcyj-msxmX0263), Chinesisch-Deutsche Mobilitätsprogamm of Chinesisch-Deutsche Zentrum für Wissenschaftsförderung (Grant No. M-0496).

# Ethical standards

The experiments comply with the current laws of China.

# Declarations

**Conflict of Interest**  The authors declare no conflict of interest.